\newcommand{\la}{\langle}
\newcommand{\ra}{\rangle}

\newcommand{\Tr}{{\rm Tr}}

\newcommand{\GeV}{{\rm GeV}}

\documentstyle[12pt]{article}

\newcommand{\beq}{\begin{eqnarray}}
\newcommand{\eeq}{\end{eqnarray}}

\begin{document}
\rightline{UTHEP-331, nucl-th/9603043}

\rightline{March, 1996}

\vspace{1cm}

\centerline{\normalsize\bf
  LIGHT VECTOR MESONS IN NUCLEAR MATTER}

\baselineskip=16pt

\vspace*{0.6cm}
\centerline{\footnotesize T. HATSUDA$^{1,2}$, H. SHIOMI$^2$ and
 H. KUWABARA$^2$}

\vspace{0.5cm}

\centerline{\it $^1$ Institute for Nuclear Theory,
 University of Washington, Seattle, WA 98195, USA}

\vspace{0.1cm}

\baselineskip=4pt

\centerline{\it $^2$ 
 Institute of Physics, University of Tsukuba,
 Tsukuba, Ibaraki 305, Japan}

\baselineskip=20pt
\vspace*{0.3cm}

\vspace*{0.9cm}
\abstract{
We summarize the current theoretical and experimental status
of the spectral changes of  vector mesons ($\rho$, $\omega$, $\phi$)
 in nuclear medium. Various approaches including QCD sum rules, effective 
 theory of hadrons and bag models  show
 decreasing of the vector meson masses in nuclear matter.
  Possibility to detect  
 the mass shift through lepton pairs
  in $\gamma-A$, $p-A$ and $A-A$ reactions are also discussed.}

\vspace{3cm}
Invited paper submitted to Prog. Theor. Phys.

\newpage

\normalsize\baselineskip=25pt
\setcounter{footnote}{0}
\renewcommand{\thefootnote}{\alph{footnote}}
\section{Introduction}

 At high temperature ($T$) and density ($\rho$),
  hadronic matter is expected to undergo  a phase  
  transition to the
  quark-gluon plasma \cite{QM95}.
 The order parameter characterizing the transition 
 is the chiral quark condensate $\langle \bar{q}q \rangle$, the 
 absolute value of which decreases as ($T$,$\rho$) increases.
  Numerical simulations of quantum chromodynamics (QCD)
 on the lattice 
  are actively pursued to determine the 
  precise nature of the transition at finite $T$ \cite{ukawa} 
 and various model calculations  have been done
 to look for the observable signature of the phase transition \cite{review}.
 
In this article, we will concentrate on one of the 
 interesting critical phenomena  associated with the 
 QCD phase transition, namely the spectral change of hadrons,
 in particular the mass shift of  light vector-mesons ($\rho$, $\omega$
 and $\phi$) in nuclear matter at zero $T$.
 The vector mesons are unique in the sense that
  they decay into lepton
 pairs ($e^+ e^-$ and $\mu^+ \mu^-$) which can be detected experimentally
 without much disturbance by  complicated hadronic interactions.
 
 In section 2, we will review the current knowledge of the 
 quark condensate in medium. In section 3, various approaches to
 calculate the vector meson masses in nuclear matter are summarized.
 Section 4 and 5 are devoted to the detailed explanation of the
 mass shift of $\rho$, $\omega$  and $\phi$ in quantum hadrodynamics, 
 an effective theory of mesons and baryons.
 Experimental possibilities to detect the spectral change 
 are discussed in section 6.  Concluding remarks are given in section 7.

\section{Quark condensates in nuclear matter}

  The dynamical breaking  of chiral symmetry has
 close similarity with the gauge-symmetry breaking in superconductors
 as has been first recognized by Nambu and Jona-Lasinio \cite{NJL}.
 Table 1  shows a brief comparison between QCD and the BCS theory
 for ``low temperature" superconductivity.

\vspace{1.0cm}

\begin{center}

\begin{tabular}{|c|c|}\hline
QCD  & BCS  \\
\hline \hline
$q-\bar{q}$ paring & $e_{\uparrow} - e_{\downarrow}$ paring \\
\hline
chiral symmetry breaking  &  gauge symmetry breaking \\
$\langle \bar{q} q \rangle \neq 0$  & 
 $\langle e_{\uparrow} e_{\downarrow} \rangle \neq 0 $ \\
\hline
quark spectrum: $E = \sqrt{p^2+M^2}$   & 
 electron spectrum: $E = \sqrt{\epsilon (p)^2+\Delta^2}$  \\
$M$ (constituent quark-mass)$\simeq 350$ MeV & 
 $\Delta$ (BCS gap) $\sim 0.01$ eV \\ \hline
Nambu-Goldstone (NG) boson = pion & NG boson is absorbed by photon
 \\ \hline 
\end{tabular} \\

\end{center}

\vspace{0.5cm}

\noindent
Table 1: Comparison between the dynamical breaking of chiral
 symmetry in QCD and the dynamical breaking of 
  gauge symmetry in the BCS theory.

\vspace{1 cm}

  The medium modification of the quark condensate
 has been calculated by various methods (lattice QCD, chiral perturbation
 theory, Nambu-Jona-Lasinio model etc). See a review \cite{review2}
 and also \cite{rev3}.
 By these studies, it turned out that
  there is one noticeable difference between the
 behavior of $\langle \bar{q} q \rangle$ at finite $T$ (with $\rho =0$)
  and that at finite
 $\rho$ (with $T=0$):  In the former case, the significant change of
 the condensate can be seen only near the critical point
 $T\sim T_c$ \cite{GL}.  On the other hand,  in the
 latter case, 
 $O(30 \%)$ change of $\langle \bar{q} q \rangle$
 could be seen even in normal nuclear-matter density. This 
 observation is based on the following formula in the
 fermi-gas approximation (independent particle approximation)\cite{DL}
\begin{eqnarray}
\label{condrho}
{\langle \bar{u}u \rangle \over \langle \bar{u}u \rangle_{0}}
  \simeq  1- {4\Sigma_{\pi N}\over f_{\pi}^2m_{\pi}^2}
 \int^{p_{_F}} {d^3p \over (2\pi)^3} {M_N \over E(p)} \ \ . 
\end{eqnarray}
Here $m_N (m_{\pi})$ is the nucleon (pion) mass, $f_{\pi}$ is the pion decay 
 constant,  $\Sigma_{\pi N} = (45 \pm 10) $MeV is the $\pi N$ sigma term,
 and $E(p) \equiv \sqrt{p^2 + M_N^2}$.  $\langle \cdot \rangle $
 and $\langle \cdot \rangle_{0}$ denote the expectation value
 in nuclear matter and that in the vacuum respectively.
 The integration for the nucleon momentum $p$ should be taken
  from 0 to the
 fermi momentum  $p_{_F}$. 
 At normal nuclear matter density ($\rho=\rho_0=0.17/{\rm fm}^3$), 
 the above formula gives (34$\pm$8)\% reduction of the
 chiral condensate from the vacuum value. 
 In Fig.1, 
${\langle \bar{u}u \rangle / \langle \bar{u}u \rangle_{0}}$ as
 well as the strangeness condensate 
$ {\langle \bar{s}s \rangle / \langle \bar{s}s \rangle_{0}}$
  are shown in the linear density approximation 
 \cite{QM91}, where the uncertainty of $\Sigma_{\pi N}$
  is considered. 
 Estimates taking into account the fermi motion and 
 the nuclear correlatons show that these
 corrections at $\rho = \rho_0$ 
 are less than the above uncertainty  \cite{Ko94}.
 
 Unfortunately, the condensate itself is not a direct observable
 and one has to look for physical quantities which are measurable and
 simultaneously  
 sensitive to the change of the condensate. 
 The masses of light vector-mesons are the leading candidates
 of such quantities.

\centerline{\bf{Fig.1}}

\section{Vector mesons in nuclear matter -- overview --}

Let's   consider $\rho$, $\omega$ and $\phi$  mesons propagating inside
 the nuclear matter.
 Adopting the same fermi-gas approximation with (\ref{condrho}) and taking
 the vector meson at rest (${\bf q}=0$),
 one can generally write the mass-squared shift  as
\begin{eqnarray} 
\label{massshiftf}
\delta m^2_{_V} \equiv m^{*2}_{_V} - m^2_{_V} = 4 \int^{p_F}
  {d^3 p \over (2 \pi)^3  }  {M_{_N} \over E(p)}  f_{VN}({\bf p}), 
\end{eqnarray}
where $f_{VN}({\bf p})$ denotes the vector-meson (V) -- nucleon (N)
 forward scattering amplitude 
 in the relativistic normalization, and $m_V^* (m_V)$ denotes the 
 vector meson mass in nuclear matter (vacuum).
 Here, we took spin-isospin average for the nucleon states
 in $f_{VN}$.
 If one can calculate $f_{VN}({\bf p})$ reasonably well in the range
 $0 < p < p_{_F}=270$ MeV
 (or $1709\ {\rm MeV} < \sqrt{s} < 1726\ {\rm MeV} $ in terms of the
 $V-N$ invariant mass), one can predict the mass shift.
 Unfortunately, this is a difficult task:
  $f_{VN}({\bf p})$ is not a  constant  
  in the above range since there are at least two 
 s-channel resonances $N(1710), N(1720)$ in the above interval
 and two nearby resonances $N(1700)$ and $\Delta(1700)$.
  They all couple to the $\rho-N$ system \cite{PDG}
 and give  variation
 of $f_{VN}({\bf p})$ as a function of $p$ in principle.
 From this reason,  one should develop
   other methods to estimate
 $\delta m_V^2$ without
 refering to the detailed form of $f_{VN}({\bf p})$.
 We will briefly review two
of such approaches
  in the following subsections, namely the QCD sum rules
 and  effective theories of hadron.

{\subsection{QCD sum rules}

 This subsection is partly based on the work in ref.\cite{HL,HLS}.
 The  QCD sum rules (QSR) for vector mesons in nuclear matter
 were first developed by Hatsuda and Lee \cite{HL}.
 In their approach,  one
 starts with  the retarded current correlation function in 
nuclear matter,
\begin{eqnarray}
\label{correlator}
\Pi_{\mu \nu} (\omega , {\bf q})
=i \int d^4x e^{iqx}  \langle {\rm R} J_{\mu}(x) J_{\nu}(0) \rangle \ \ ,
\end{eqnarray}
where  $q^\mu \equiv (\omega , {\bf q})$ and 
 ${\rm R} J_{\mu}(x) J_{\nu}(0) \equiv 
  \theta(x^0) [J_{\mu}(x), J_{\nu}(0)]$ with the source
currents $J_\mu$ defined as $J_\mu^{\rho,\omega} 
={1\over2}(\bar{u}\gamma_\mu u \mp \bar{d}\gamma_\mu d)$ 
($- (+)$ is for the $\rho^0 (\omega)$-meson) and
 $J_{\mu}^{\phi} = \ \bar{s} \gamma_{\mu} s$.
  Although there are two independent
invariants in medium (transverse and longitudinal polarization),
 they coincide  in the limit  ${\bf q} \rightarrow 0$
 and reduce to
 $\Pi_{\mu \mu}/(-3\omega^2) \equiv \Pi$.
 $\Pi $ satisfies the following dispersion
relation,
\begin{eqnarray}
\label{dispersion2}
{\rm Re} \Pi (\omega^2) =
 {1 \over \pi} {\rm P} \int_0^{\infty} du^2
{ {\rm Im} \Pi(u) \over u^2-\omega^2} + ({\rm subtraction}). 
\end{eqnarray}
In QSR, the spectral density ${\rm Im} \Pi$
 is  modeled with several 
phenomenological parameters, while  ${\rm Re} \Pi$
  is calculated using the 
operator product expansion (OPE).
 The phenomenological parameters are then  extracted  
by matching the left and right hand side of (\ref{dispersion2})
 in the asymptotic region $\omega^2 \rightarrow - \infty $.
 The density dependence in the OPE side is solely determined by the
 density dependent condensates which are evaluated from
 low energy theorems or from the parton distribution of the 
  nucleon \cite{HL}.

 In the medium,  we have three kinds of structure in the spectral density:
 the resonance poles,
  the continuum and the Landau damping contribution.
 For ${\bf q} \rightarrow 0$, the last contribution 
 is calculable {\em exactly} 
 and behaves like  a pole at $\omega^2=0$ \cite{HL,BS}.
  In total,  the hadronic spectral function looks as
\begin{eqnarray}
\label{phen}
 8 \pi {\rm Im} \Pi(u > 0^-) & = & 
 \delta(u^2) \rho_{sc}+F^* \delta(u^2-m_V^{2*})
+(1+\frac{\alpha_s}{\pi})\theta(u^2-S_0^*) \\ \nonumber
 & \equiv & \rho_{had.}(u^2) ,
\end{eqnarray}
with $\rho_{sc} = 2 \pi^2 \rho /\sqrt{p_F^2 + M_N^2} \simeq
 2 \pi^2 \rho /M_N$.  $m_V^*$, $F^*$ and $S_0^*$ 
 are the three phenomenological
 parameters in nuclear matter to be determined by the sum rules.

Matching the OPE side 
 and the phenomenological side via the dispersion relation 
in the asymptotic region $\omega^2 \rightarrow - \infty$,
 we can 
  relate the resonance  parameters to the density dependent condensates.  
 There are two major procedures for this matching, namely 
 the Borel sum rules (BSR) \cite{SVZ} and the
  finite energy sum rules (FESR) \cite{PIV},
 which can be summarized as
\begin{eqnarray}
\label{sumrules}
\int_0^{\infty}& ds\ W(s)& \ [\rho_{had.}(s) - \rho_{_{OPE}}(s) ]  =0 ,\\ 
& W(s) & = \left\{ \begin{array}{ll}
                    s^n \ 
 \theta(S_0 -s) & \ \ \ \ \ \ \ \ \  ({\rm FESR}), \\ 
\nonumber
                    e^{-s/M^2} & \ \ \ \ \ \ \ \ \ ({\rm BSR}).
                   \end{array}     \right.
\end{eqnarray}
Here the spectral function $\rho_{had.}(s)$ stands for eq.(\ref{phen}).
 $\rho_{_{OPE}}(s)$ is a hypothetical imaginary part of
 $\Pi$ obtained from OPE.

 To make  quantitative analyses of spectral 
 parameters, the stability analysis
 based on the  Borel transform is more suitable than FESR.
 Since the Borel mass $M$ is a fictitious parameter introduced 
 in the sum rule, the physical quantities should be insensitive to
 the change of $M$ within a Borel interval $M_{\min} < M < M_{\max}$;
 namely the principle of minimum sensitivity (PMS) \cite{ste} is used.
 One can accomplish this insensitivity by choosing
 $S_0^*$ suitably at given density.
 In Fig. 2, the  Borel curves for the $\rho (\omega)$ meson
 for three different values of baryon density are shown with $S_0^*$
 chosen to make the Borel curve as flat as possible in the interval  
 $ 0.41 {\rm GeV}^2 < M^2 < 1.30 {\rm GeV}^2$. The upper (lower)
 bound of the Borel interval is determined so that the power (continuum)
 correction after the Borel tranform does not exceed 30 $\%$ of 
 the lowest order term in OPE.  
 
\centerline{\bf{Fig.2}}

  By making a linear fit of the result,  one obtains \cite{HL,HLS}
\begin{eqnarray}
\label{mass-shift}
{m_{\rho,\omega}^* \over m_{\rho,\omega}} & = & 1- (0.16 \pm 0.06) 
{\rho \over \rho_0}, \\
\label{threshold-shift}
\sqrt{{S_0^* \over S_0}} & = & 1- (0.15 \pm 0.05)
 {\rho \over \rho_0}, \\
\label{F-shift}
{F^* \over F} & = & 1- (0.24 \pm 0.07) {\rho \over \rho_0}, 
\end{eqnarray}
and 
\begin{eqnarray}
\label{mass-shift2}
{m_{\phi}^* \over m_{\phi}}  =  1- (0.15 \pm 0.05)\  y\  
{\rho \over \rho_0},
\end{eqnarray}
where $y$ is the OZI breaking parameter in QCD defined
 as $y=2\la \bar{s}s \ra_N/ \la \bar{u}u + \bar{d} d \ra_N$
 with $\la \cdot \ra_N$ being the nucleon matrix element.
 $y$ takes the value
 $0.1 - 0.2$ \cite{HL}.  
  The decrease in eqs. (\ref{mass-shift},\ref{mass-shift2}) 
 is dictated by the density dependent
  condensates $\langle \bar{q} q \rangle$,  $\langle (\bar{q} q)^2 \rangle$
 and  $\langle \bar{q} \gamma_{\mu} D_{\nu} q \rangle$.
 The errors in the above formulas are originating from the uncertainties
 of the density dependence of the these condensates.
 The contribution of the
  quark-gluon mixed operator with twist 4, \cite{HLS} which may
   possibly weaken the mass shift, is neglected in the above.
   Shown in Fig.3 is the mass shift given in 
 eqs. (\ref{mass-shift},\ref{mass-shift2}) with possible 
 theoretical uncertainties.

\centerline{\bf{Fig.3}}

 Some sophistications of the QSR analyses by Hatsuda and Lee
 have been done later by several authors.

\noindent
(i) Asakawa and Ko have introduced  a more realistic
 spectral function  than (\ref{phen})  by taking into account
 the width of the $\rho$-meson and the effect of  $\pi - N - \Delta -\rho$
 dynamics \cite{ASKO}.  By doing the similar QSR analysis as above,
 they found that the negative mass shift 
  occurs also  in this realistic case.

\noindent
(ii) Monte Calro based error analysis was applied to the 
 Borel sum rule by Jin and Leinweber \cite{JL} instead of the
 Borel stability or PMS analysis employed in \cite{HL,HLS}.
 They found 
 $ m_{\rho,\omega}^*/ m_{\rho,\omega} =  1- (0.22 \pm 0.08)  
(\rho /\rho_0)$ and 
 $m_{\phi}^* /m_{\phi} =  1- (0.01 \pm 0.01) 
(\rho / \rho_0)$, which are consistent with
 eqs. (\ref{mass-shift},\ref{mass-shift2}) within 
 the error bars.

\noindent
(iii) Koike analysed an {\em effective} scattering amplitude $\bar{f}_{VN}$
 defined as   $\delta m^2_{_V} \equiv {\bar f}_{VN} \cdot \rho $ 
 using the QSR in the vacuum \cite{Koike}.  Although his original calculation
 predicting ${\bar f}_{VN} > 0$ is in error as was pointed out in
 ref.\cite{HLS,JL}, revised calculation gives a consistent result with
 eqs. (\ref{mass-shift},\ref{mass-shift2}) within the error bars \cite{Koike2}.
 Note here that ${\bar f}_{VN}$ does not have 
  direct relation to the scattering length at zero
 momentum $f_{VN}(0)$.

\subsection{Effective theories}

 There have been  many attempts so far to calculate the 
 spectral change of the vector mesons using effective theories of QCD.
 The first attempt by Chin \cite{chin}
 using the quantum hadrodynamics (QHD) 
 shows increasing $\omega$-meson mass
 in medium due to a process  analogous to the Compton scattering;
\begin{eqnarray}
 \omega + N 
  \rightarrow \omega + N.
\end{eqnarray}
  For the $\rho$-meson, similar but more sophisticated calculations
 taking into account  $\Delta$-resonance and in-medium pion
 show a slight increase of the $\rho$-meson mass \cite{herman}.
  In these calculations, only the 
  polarization of the Fermi sea (the particle-hole excitations)
 was considered. Also their predictions are
 different from the general assertion by Brown and Rho claiming
 that all the hadron masses except for pion should decrease \cite{BR}.

  On the other hand, Saito, Maruyama and Soutome \cite{SS} and 
  Kurasawa and Suzuki \cite{KS}
   have realized that 
  the mass of the $\omega$-meson is affected substantially
 by the vacuum polarization of the nucleon in medium 
\begin{eqnarray}
 \omega \rightarrow N^* \bar{N}^* \rightarrow \omega,
\end{eqnarray}
 where  $N^*$ is the nucleon
  in nuclear matter which has smaller effective mass than that in the vacuum.
  They show  that the vacuum polarization  dominates over the 
 Fermi-sea polarization in QHD  and leads decreasing
 vector meson mass. This conclusion was later confirmed by several
 groups \cite{TBA,CJ,PW} and was generalized for the $\rho$ and $\phi$ mesons
  by the present authors \cite{HS,kuwa} which will be
 discussed in more details in section 4 and 5.
  Jaminon and Ripka has also reached a similar conclusion
  by using a model of vector mesons coupled to  constituent quarks
  \cite{JR}.
 
  Saito and Thomas have examined a rather different 
 but comprehensive model (bag model combined with QHD)
 and 
 found  decreasing vector-meson masses \cite{ST};
 $ m_{\rho,\omega}^* / m_{\rho,\omega}  \sim   1- 0.09  
(\rho / \rho_0)$.
  The spectral shift 
 of the quarks inside the bag induced by the existence of 
 nuclear medium plays a key role in this approach. 

   Basic idea common in the approaches predicting the
 decreasing mass may be summarized as follows.
 In nuclear matter, scalar ($\sigma$) and vector ($\omega$)
 mean-fields are induced by the  nucleon sources.
  These mean-fields give back-reactions
 to the nucleon propagation in nuclear matter and modify
  its self-energy. This is an origin of the effective nucleon mass
 $M_N^* < M_N$ in  the relativistic
 models for nuclear matter.
  The same mean-fields should also affect the propagation of 
  vector mesons in nuclear medium.
 In QSR, the quark condensates act on the quark propagator
 as density dependent mean-fields.
  In QHD, the coupling of the mean-field with the vector
 mesons are taken into account through the short distant
  nucleon
  loop with the effective mass $M_N^*$.
  In the bag-model, the mean fields outside the bag
 acts on quarks confined in the 
 bag and change their energy spectrum. 
 
 Let us show here that one can understand the 
 negative mass shift of the vector mesons in a simple
 and intuitive  way in the context of QHD.
 More quantitative discussion will be given in the later section.
  After renormalizing infinities in the vacuum loop,
  the density-dependent part of the
 Dirac-sea polarization
 to the vector-meson propagator is approximately  written as 
\begin{eqnarray}
\label{Zfactor}
D(q) \simeq {1 \over Z^{-1} q^2 - m_V^2} = {Z \over q^2 - Z m_V^2} ,
\end{eqnarray}
where $Z$ being the finite wave-function renormalization constant
 in medium.  The pole position is thus obtained as $m_V^* = \sqrt{Z} m_V$.
 Because of the current conservation, only the
 wave function part of the propagator is modified in medium.
 Since the effective mass of the nucleon decreases in medium
 ($M_N^*/M_N < 1$), physical 
 vector mesons have more probability to be in
 virtual baryon$-$ anti-baryon
  pairs compared to that in the vacuum.  This means
 $Z < 1$, which leads to $m_V^* /m_V \equiv Z  < 1$ \cite{PW,HS}.
\begin{eqnarray}
M_N^*/M_N < 1 \ \ \  \rightarrow \ \ \  Z < 1 \ \ \  \rightarrow \ \ \ 
  m_V^*/m_V \equiv Z   < 1 \ \ \ . 
\end{eqnarray}

\section{$\rho$ and $\omega$ mesons in quantum hadrodynamics}

\subsection{Nucleon at finite density}
\label{nm}
 Before discussing 
  the vector meson masses in medium, let's
   make a brief review of the effective nucleon mass at finite density
  in the quantum hadrodynamics (QHD) on the basis
   of ref.\cite{wal74,chin,SW86}.
   The lagrangian of QHD is written as
\beq
{\cal L}&= &  \bar{\psi}
 [\gamma_\mu(i \partial^\mu-g_v V^\mu)-(M_N -g_s S)
 ] \psi  \nonumber \\
&+& {1 \over 2} (\partial_\mu S \partial^\mu S-m_s^2 S^2)
-{1 \over 4} F_{\mu \nu}F^{\mu \nu} +{1 \over 2} m_v^2 V_\mu V^\mu
+ {\cal L_{CT}} \ \  ,
\eeq
with
\beq
F_{\mu \nu}=\partial_\mu V_\nu-\partial_\nu V_\mu \ \ ,
\eeq
where $\psi(x), S(x)$ and $V(x)$ are
 nucleon, $\sigma$ meson and $\omega$ meson field,  respectively.
 $\sigma (\omega)$ field is 
coupled to nucleon current with the coupling constant
 $g_s (g_v)$.
${\cal L_{CT}}$ is the counter term added to the original lagrangian 
in order to get the finite physical  quantities.

Let's define the nucleon propagator in uniform nuclear matter,
\beq
iG_{\alpha \beta}(x, y)=\la {\rm T}
 \psi_\alpha (x) \bar{\psi}_\beta (y) \ra \ \ .
\eeq
 The free nucleon propagator in momentum space is expressed as
\beq
\label{HA}
G^0(p)&=&(\gamma \cdot p +M_N)\left \{ {1 \over p^2-M_N^2 + i\epsilon}
+{ i \pi \over E(p)} \delta(p_0-E(p)) \theta(p_F- |{\bf p}|) \right \},
 \nonumber \\
&\equiv& G_F^0(p)+G_D^0(p),
\eeq
where $E(p)=\sqrt{{ \bf p }^2+ M_N^2}$ and $p_F$ is the fermi momentum.
 In the relativistic Hartree approximation (RHA) \cite{SW86,chin}, 
  the full nucleon propagator reads 
\beq
\label{RHA} 
G^H(p)&=& G^0(p)+G^0(p) \Sigma^H  G^H(p),
\eeq
where $H$ denotes the Hartree approximation.
 A schematic diagram for $G^H(p)$  is given in Fig.4.

\centerline{{\bf Fig.4}}

$\Sigma^H$ in Fig.4 can be written as
\beq
\Sigma^H  &=&\Sigma_s- \gamma_\mu \Sigma_v^\mu, \\
\label{scc}
\Sigma_s &=&i {g_s^2 \over m_s^2} \int {d^4 p \over (2\pi)^4}
 \Tr [G^H(p)],  \\
\label{svv}
\Sigma_v^\mu &=&i {g_v^2 \over m_v^2} \int {d^4 p \over (2\pi)^4}
 \Tr [ \gamma^\mu G^H(p)],  
\eeq
where $\Sigma_s (\Sigma_v^\mu)$ is scaler (vector) self-energy. 
 Formal solution of  eq.(\ref{RHA}) reads
\beq
[G^H(p)]^{-1}&=&\gamma \cdot p-M_N - \Sigma^H  \nonumber \\
            &=& \gamma \cdot( p+ \Sigma_v)-(M_N+ \Sigma_s),
\eeq
or equivalently
\beq
\label{RHAS}
G^H(p)&=&(\gamma \cdot \bar{p} +M_N^*)\left \{ {1 \over \bar{p}^2-M_N^{*2}
 + i\epsilon}
+{ i \pi \over E^*(p)} \delta(\bar{p}_0-E^*(p)) \theta(p_F- |{\bf p}|) \right \},
 \nonumber \\
&\equiv& G_F^H(p)+G_D^H(p),
\eeq
where $E^*(p)=\sqrt{{\bf p }^2 + M_N^{*2}}, \bar{p}=p+\Sigma_v$
 and $M^*_N=M_N +\Sigma_s$.
Eq. (\ref{RHAS}) implies that
the interacting propagator    
  can be separated 
into two parts:
 $G_F^H(p)$ which  has the same form as the free nucleon  propagator  
 with the effective nucleon mass $M_N^*$, and 
  $G_D^H(p)$ which depends explicitly  on  the fermi momentum  
in  nuclear matter.
The advantages  of this separation will be discussed later. 

Next we turn to discuss self-energy $\Sigma^H$ in which  
 $\Sigma_s$ is related to the nucleon effective mass.        
For the vector self-energy,  insertion of eq.(\ref{RHAS}) into 
eq.(\ref{svv}) gives, 
\beq
\Sigma_v^\mu= 8 i {g_v^2 \over m_v^2} \int {d^4 p \over (2\pi^4)}
{ \bar{p}^\mu \over \bar{p}^2-M^{*2}_N  + i \epsilon}
-{g_v^2 \over m_v^2}\delta^{\mu 0} \rho_B, 
\eeq 
where $\rho_B={\gamma \over 6 \pi^2 }p_F^3$  is called baryon density
 with  $\gamma=4 $ for nuclear matter.
      By shifting the integration  variables from $p $ to  $\bar{p} $, 
      the first term on the r.h.s.  vanishes, 
        namely $\Sigma_v^\mu$ has no  divergent term.
In the case of  scalar  self-energy, however,  one gets  
\beq
\Sigma_s=i {g_s^2 \over m_s^2} \int { d^4 p \over (2 \pi)^4}
{8 M^*_N  \over {\bar{p}^2-M^{*2}_N + i \epsilon}}
-{g_s^2 \over m_s^2} { \gamma \over (2\pi)^3}
\int^{p_F} d^3 p {M^*_N \over ( {\bf p }^2+ M^{*2}_N )^{1/2}},
\eeq  
where   the first term on the r.h.s.            
 is divergent.
To single out this divergence, we extract the infinite part 
using dimensional regularization:
\beq
\Sigma_s^{inf} 
&=&-{g_s^2 \over m_s^2}{\Gamma(2-{n \over 2} ) \over 2 \pi^2} M^{*3}_N
\nonumber  \\
&=&-{g_s^2 \over m_s^2}{\Gamma(2-{n \over 2} ) \over 2 \pi^2}
(M^3_N + 3 M^2_N \Sigma_s +3 M_N  \Sigma_s^2 +\Sigma_s^3),
\label{diver}
\eeq
where $\Sigma_s^{\inf}$ denotes 
 the infinite parts and the diagrammatic
 illustrations are depicted in Fig.5.         

\centerline{{\bf Fig. 5}}

These divergent terms are removed by defining  the following 
 ${\cal L_{CT}}$;   
\beq
{\cal L_{CT}}
= \alpha_1 S + {\alpha_2 \over 2!} S^2 
+ {\alpha_3 \over 3!}  S^3 
+ {\alpha_4 \over 4!} S^4,  
\eeq
 which
  yields new contributions,  $\Sigma^{CT}_s$, 
 to the scalar self-energy: 
\beq
\Sigma^{CT}_s=\sum_{n=0}^3 {1 \over n!}
({-g_s \over m_s^2})({-\Sigma_s \over g_s})^n \alpha_{n+1}.
\eeq   
The corresponding Feynman diagrams are depicted in Fig.6.

\centerline{{\bf Fig. 6}}

 $\alpha_1-\alpha_4$ are determined so as to 
  cancel precisely the  divergent parts in (\ref{diver}):
\beq
\alpha_n =(-i)(-g_s)^n (n-1)! \int { d^4 p \over (2 \pi)^4} \Tr[G_F^0(p)^n]. 
%
%
%
\eeq

  Note that the finite parts of $\alpha_{1-4}$ are also fixed in the
  above conditions \cite{chin}, which may or may not be 
   justified and must be checked using experimental/empirical inputs.
 Examining this point is not a main theme of this article, and we
  simply take the above procedure to calculate the density dependent
   part of $\Sigma^H_s$: 
\beq
\Sigma^H_s &=& 
-{g_s^2 \over m_s^2} \rho_s
+ {g_s ^2 \over m_s^2}{ 1\over \pi^2}
[ M^{*3}_N \log (M^*_N/M_N) \nonumber  \\ 
& &  -M^2_N(M^*_N-M_N)-{5 \over 2}M_N(M^*_N-M_N)^2
-{11 \over 6}(M^*_N-M_N)^3 ] \nonumber \\
&\equiv& M^*_N-M_N,
\eeq
where
\beq
\label{scaden}
 \rho_s={M^*_N \over \pi^2} \left \{ p_F E^*(p_F)-M^{*2}_N \log 
\left | p_F+E^*(p_F) \over M^*_N \right | \right \}.
\eeq
This expresses	 the 
  effective nucleon mass including the vacuum polarization.
The density dependence of the effective nucleon mass is shown in 
 Fig.9.  $g_s, g_v$ and $m_s$ are chosen to satisfy  
 the saturation density for nuclear matter ($-$15.75 MeV)
 at nuclear matter density $p_F=1.30$ $ {\rm fm}^{-1}$:
  $g_s= 7.37, g_v=10.1$ and $ m_s=458 {\rm MeV}$ \cite{PW}. 
Hence  we find that effective nucleon mass has
 the reduction, $M^*_N/M_N=0.730$,  at
nuclear matter density.
 As will be shown in section.\ref{vmfd}, 
the reduction plays a crucial role in studying   vector meson
 masses at finite density.  

\subsection{$\rho$ and $\omega$ mesons at finite
 density}

 This subsection is partly based
 on the work in ref.\cite{HS}.

\label{vmfd}
\subsubsection{Effective lagrangian}
Let's start with an  interaction lagrangian of $\rho,\omega$ with
 the nucleon:
\beq
\label{interaction}
L_{int} =g_v \left [ \overline{{\psi}} \gamma_\mu \tau^a {\psi} 
-\frac{\kappa_v}{2M_N} \overline{{\psi}} \sigma_{\mu \nu} \tau^a {\psi}
\partial^\nu \right ] V^\mu_a , \hspace{0.5cm} v=\{\rho, \omega\}\ \ ,
\eeq
where $a$ runs from 0 through 3, $V_0$ ($V_{1-3}$) corresponds to
 the $\omega$ ($\rho$) field, $\tau^a$ is the isospin matrix 
 with $\tau^0$=1, and $M_N$ is the nucleon mass.
 The numerical values of the coupling constants
  ($g_{v}$, $\kappa_{v}$)  will be given in sec. 4.2.4.

 In nuclear medium,  meson propagator is defined as
\beq
D^{\mu \nu}(x,y)=\la  {\rm T} V^\mu(x) V^\nu(y)  \ra.
\eeq
 By using the free vector-meson propagator
\beq
 D_0^{\mu \nu} (q)={ g_{\mu \nu} \over q^2-m^2 +i \epsilon}
+{1 -\lambda \over \lambda}
 {q_\mu q_\nu  \over (q^2-m^2 / \lambda + i \epsilon)(q^2-m^2 +i \epsilon)}, 
\label{stk}
\eeq
 with $\lambda$ being a gauge parameter in the Steukelberg
  formalism \cite{IZ},
 the full propagator can be written as
\beq
D^{\mu \nu}(q)=D^{\mu \nu}_0(q)
 + D^{\mu \lambda}_0 (q) \Pi_{\lambda \sigma}(q) D^{\sigma \nu}(q),
\eeq
with the self-energy $\Pi_{\mu \nu}$.  See Fig.7.a.

\centerline{{\bf Fig.7}}

In the one-loop level, the density dependent part of the 
 self-energy  comes only from  the nucleon-loop (Fig.7.b):
\beq
\label{polarization}
\Pi_{\mu \nu}^{ab}(q)=-\frac{i}{(2\pi)^4}\int d^4k 
{\rm Tr}[\Gamma_\mu^a G^H(k+q) \tilde{\Gamma}_\nu^b G^H (k)]  \ \ ,
\eeq
where $(a,b)$ are the isospin indices and  we have used $G^H(p)$
 defined in section \ref{nm}.
For the vertices, we make use of   
\beq
\label{vertex}
\Gamma_\mu^a=g_v [\gamma_\mu \tau^a - 
\frac{\kappa_v}{2M_N}\sigma_{\mu \lambda} iq^\lambda \tau^a], \hspace{1cm}
\tilde{\Gamma}_\nu^b=g_v [\gamma_\nu \tau^b + 
\frac{\kappa_v}{2M_N}\sigma_{\nu \lambda} iq^\lambda \tau^b]\ \ .
\eeq
where the relative sign of the 
 tensor part in $\Gamma_\mu$ and that in $\tilde{ \Gamma}_\nu$
 is opposite to that in ref.\cite{shakin}.   
The self interaction
 of the $\rho$ meson gives density dependence only from two or higher 
  loops.
  The coupling of $\rho$
 with in-medium pions analyzed in \cite{herman}
 is also  the higher loop effect and will not be  considered in this paper.

 $\Pi_{\mu \nu} $  in (\ref{polarization}) is composed
 of two parts $\Pi_{\mu \nu} = \Pi^{0F}_{\mu \nu} + \Pi^D_{\mu \nu}$:
 the first term corresponds to the
 fluctuation of the Dirac sea of the nucleons with mass $M_N^*$,
  while the second term
  gives  the fluctuation of the 
  Fermi sea + the Pauli blocking. 
The advantage of taking this separation is that
  $\Pi_F^0$ and $\Pi_D$ satisfy  
 current conservation separately, i.e.,   $p^\mu \Pi_{\mu \nu}^{0F}(p)
=p^\mu \Pi_{\mu \nu}^D=0$ \cite{DF90}.
 If one adopts a  separation without current conservation, 
 the corresponding self-energy induces spurious results
  \cite{DF90}. 
 $\Pi^{0F}_{\mu \nu}$
  generally  has divergences to be subtracted.  We 
 will show our subtraction procedure in section \ref{dcs}
  and define $\Pi^{F}_{\mu \nu}$
 as the subtracted polarization.

The vector meson propagator in the medium has a general form
\beq
\label{mesonpropagator}
D^{\mu \nu} = {-P_L^{\mu \nu} \over q^2-m^2 + \Pi_L}
              + {-P_T^{\mu \nu} \over q^2-m^2 + \Pi_T}, 
\eeq
where we have suppressed isospin indices $(a,b)$ and
 $m$ denotes the $\rho$ or $\omega$ mass in the vacuum.
  $P_{T}^{\mu \nu}$ ($P_{L}^{\mu \nu}$) is the projection 
operator
 to the
 transverse (longitudinal) direction to
  ${\bf q}$: 
\beq
 P_T^{\mu \nu}& =
&  g^{\mu i} (g_{ij} + q_i q_j/{\bf q}^2)g^{\nu j}, \ \ \ \ 
 P_L^{\mu \nu} = e^{\mu}e^{\nu} 
 \\
 \mbox{with} & & 
 e^{\mu}= {i \over \sqrt{q^2}}(|{\bf q}|,q_0{\bf q}/|{\bf q}|).
\eeq
  $\Pi_{T,L}$ is related to $\Pi_{\mu \nu}$ as
\beq
\Pi_L = - (q^2/{\bf q}^2) \Pi_{00}, \ \ \ \  
\Pi_T = (\Pi_l^l + (q_0^2/{\bf q}^2) \Pi_{00})/2.
\eeq
In this  paper we will focus on $\Pi_{T,L}$    
 in the time like region  with ${\bf q}=0$ where
 $\Pi_L$ is exactly equal to $\Pi_T$.  
To  obtain (\ref{mesonpropagator}), we adopt (\ref{stk}) with
  $\lambda \rightarrow \infty$   as a free propagator of the
 massive vector mesons.
\subsubsection{Dirac sea effect and subtraction procedure}
\label{dcs} 
The interaction (\ref{interaction}) is not  a renormalizable one 
  in the conventional sense. This implies
 that the model contains 
 infinite series of the higher dimensional operators 
 which play a role to cancel the divergences
 emerging from the loops of the lower dimensional operators \cite{WEIN,pol}.
  Instead of developing a systematic subtraction
 procedure, we
 will take a phenomenological way to extract $\Pi_{\mu \nu}^F$
 from $\Pi^{0F}_{\mu \nu}$.
  First of all, we will normalize the propagator at zero density
  as $1/(q^2 - m^2)$, i.e. subtract away both the divergent and finite parts
 from $\Pi_{\mu \nu}^{0F}$.
 This corresponds to a set of the renormalization conditions
 at zero density,
\beq
\left. {\partial^n \Pi^F(q^2) \over
 \partial (q^2)^n } \right| _{q^2=m^2} = 0
 \ \ \  (n=0,1,2, \cdot \cdot \cdot ,\infty ) .
\eeq
 A straightforward generalization of the above conditions 
 to finite density  
 reads 
\beq
 \left. {\partial^n \Pi^F(q^2) \over \partial
 (q^2)^n } \right| _{M^*_N \rightarrow M_N, q^2=m^2} = 0 
  \ \ \ (n=0,1,2, \cdot \cdot \cdot , \infty ).
\eeq
   This together with a requirement that
 the higher dimensional counter terms are the local polynomials
  allow one to single out the density dependent part 
 of $\Pi_{\mu \nu}^{0F}$ uniquely.

 Our renormalization condition
 is different from the simple scheme 
 $\Pi_{\mu \nu}^F = \Pi_{\mu \nu}^{0F} - \Pi_{\mu \nu}^{0F}\mid_{M^*_N=M_N}$.
 This simple scheme
  cannot
remove the divergences such as $M^*_N/\epsilon$ and $(M^*_N)^2/\epsilon$
 $(\epsilon \rightarrow 0$) induced by the  $\rho NN$ tensor coupling.
  These divergences require the counter 
 terms such as $ S^n F_{\mu \nu}
F^{\mu \nu}$ ($S$ being the scalar field and  $F_{\mu \nu}$ being the 
 field-strength for the vector fields) which are   
 allowed in our scheme  but not in the simple scheme.
  Although our  procedure is physically plausible,
 it is still ``a'' way to subtract the divergences among many other
 possibilities.  
 For the $\omega$ meson,  our procedure
  is equivalent to that in \cite{CJ}.
 For $m_{\omega}^*$, we have checked that
  the different subtraction procedures in \cite{KS,PW} 
   does not cause more than $3 \%$ differences
   from ours at $\rho = \rho_0$. This is shown in Fig. 8.
 
\centerline{{\bf Fig.8}}

 Using the dimensional reguralization and the above subtraction procedure,
  one obtains the following $\Pi_{\mu \nu}^F$  for the $\rho$ and $\omega$
   mesons. 
\beq
\label{result}
\Pi^{ ab F }_{\mu \nu} 
&=&  \delta^{ab} 
( q_{\mu}q_{\nu}/q^2 - g_{\mu \nu})
 (\Pi_v^F + \Pi_{v, t}^F+\Pi_t^F) \nonumber \\
\Pi_v^F&=& \frac{g_v^2}{\pi^2} q^2 
\int^1_0 dx \ x(1-x)\log \left \{ \frac{{M^*_N}^2-q^2x(1-x)}{M^2_N-q^2x(1-x)}
 \right \},\\
\Pi_{v,t}^F &=& (\frac{g_{v}^2\kappa_v}{2M})\frac{M^*_N q^2}{\pi^2} 
\int^1_0 dx \log \left \{\frac{{M^*_N}^2-q^2x(1-x)}{M^2_N
-q^2x(1-x)}\right \},\\
\label{result1}
\Pi_t^F &=& (\frac{g_{v} \kappa_v }{2M})^2 \frac{q^2}{2\pi^2}
\int^1_0 dx \{{M^*_N}^2 +q^2 x(1-x) \}
\log \left \{ \frac{{M^*_N}^2-q^2x(1-x)}{M^2_N -q^2x(1-x)}\right \},
\eeq
where the integrals take analytic forms:
\beq
& &\int^1_0 dx \ x(1-x)\log \left \{ \frac{{M^*_N}^2-q^2x(1-x)}{M^2_N
-q^2x(1-x)}
 \right \}
  \nonumber \\
& & \ \ \ 
 = {1 \over 3}  \log (M^*_N/M_N)-{2 \over 3} \left ({M^{*2}_N \over q^2} 
 -{M^2_N \over q^2} \right )  \nonumber \\
 & & \ \ \ +  
 {A^* \over 3} \left ({2 M^{*2}_N \over q^2} +1 \right )
 \tan^{-1}(1/A^*)- 
{A \over 3} \left ({2 M^2_N \over q^2} +1 \right )
 \tan^{-1}(1/A) ,
 \\
& &\int^1_0 dx \ \log \left \{ \frac{{M^*_N}^2-q^2x(1-x)}{M^2_N-q^2x(1-x)}
 \right \} \nonumber   \\
 & & \ \ \ =
2 \log(M^*_N / M_N) +2 A^* \tan^{-1}(1/A^*)-
    2 A \tan^{-1}(1/A),    
\eeq
where $ A^*=\left( |{4 M^{*2}_N \over q^2}-1| \right )^{1/2}, 
A=\left( |{4 M^2_N \over q^2}-1| \right )^{1/2}$, and $4M^{*2}_N > q^2$
 is assumed.

\subsubsection{ Fermi sea effect}
 Another contribution  to $\Pi_{\mu \nu}$ comes from 
  the Fermi sea
   and the Pauli blocking effects, which is expressed  as
\beq
\label{DDp}
\Pi_{\mu \nu}^{abD}(q) &=&
-{i \over (2 \pi)^4} \int d^4 k
\Tr [\Gamma^a_\mu G_F(k+q) \tilde{\Gamma}^b_\nu G_D(k) +
 (F \leftrightarrow D)] \\
&=&
 \delta^{ab}(\Pi^D_v+ \Pi^D_{v,t}+\Pi^D_t)_{\mu \nu}.
\eeq
 Substituting eq.(\ref{RHAS}) into eq.(\ref{DDp}), 
 one obtains the following  form:
\beq
(\Pi^D_v)_{\mu \nu} &=& g_v^2 {\bar \Pi}_{\mu \nu}(q), \\
(\Pi^D_{v,t})_{\mu \nu}
 &=& Q_{\mu \nu}  
 \left(\frac{g_{v}^2 \kappa_v}{2M_N}\right)4M^*_N  q^2 I_0(q),\\
(\Pi^D_t)_{\mu \nu} &=& \left (\frac{g_{v} \kappa_v }{2M_N}\right)^2 q^2  
 \left [ - \bar{\Pi}_{\mu \nu}(q)
  + Q_{\mu \nu} \left \{(4{M^*_N}^2+q^2)I_0(q)+ {\rho_s \over {M^*_N}}
  \right \} \right ], 
\eeq
where $Q_{\mu \nu}=(q_\mu q_\nu/q^2 - g_{\mu \nu})$ and $\rho_s$ has been
 defined in (\ref{scaden}), and 
\beq
\bar{\Pi}_{\mu \nu}(q) 
&=&
{2  \over \pi^3} \int d^3 k { \theta(p_F-{\bf k}) \over E^*(k)}
{ 1 \over q^4 -4(k \cdot q)^2},  \nonumber \\
& \times &
 \left \{ (k \cdot q)^2 \left [ g_{\mu \nu}-{q_\mu  q_\nu \over  q^2 } \right]
+ q^2 
\left [ k_\mu- { k \cdot q \over q^2} q_\mu \right]
\left [ k_\nu- { k \cdot q \over q^2} q_\nu \right] \right \}, \\
I_n(q)&=&J_n(q)+J_n(-q), \\
 J_n(q)&
= &{-2 \over (2 \pi)^3} \int {d^3 k \over E^*(k)}
 { \theta(p_F- {\bf k}) \  (2 E^*(k) +q_0)^n
 \over q^2 + 2 q\cdot k + i\epsilon} \left |_{k_0=E^*(k)} \right. . 
\eeq
Here we have shown
 only the real parts of $\Pi_D$ relevant to study the pole position at
 ${\bf q} =0$.

In the limit of ${\bf q}=0$,  
 $ I_0(q)$ and $\bar{\Pi}_{T,L}(q)$ defined from $\bar{\Pi}_{\mu \nu}(q)$
 read 
\beq
& & \bar{\Pi}_T(q_0, {\bf q}=0)
= \bar{\Pi}_L(q_0, {\bf q}=0)
   \\
&=&
{2   \over 3 \pi^2}
\left \{ p_F E^*(p_F) +{ q_0^2 \over 2}
 \log \left | { E^*(p_F)+p_F \over M^*_N} \right|
- \left (M^{*2}_N +{q_0^2 \over 2} \right )
 A^* \tan^{-1} \left ({p_F\over E^*(p_F) A^*} \right ) \right \},  \nonumber \\
& &I_0(q_0, {\bf q}=0 )= { 1\over 4 \pi^2 }
\left \{ 2 \log \left |{p_F + E^*(p_F) \over M^*_N} \right |
 -2 A^* \tan^{-1}\left ({p_F \over E^*(p_F) A^*} \right ) \right \} ,
\eeq
where $ A^*=\left( |{4 M^{*2}_N \over q_0^2}-1| \right )^{1/2}$
 and $4 M^{*2} > q_0^2$ is assumed.  

 The mixing  
 of the $\omega$ meson and the  $\sigma$ meson does not contribute 
to the propagator
 as far as ${\bf q} =0$ and $q_0^2 > 0$. 
This is because
 the meson self-energy
    relevant for  the $\omega -\sigma$ mixing vanishes exactly
  in the time-like region with ${\bf q}=0$   
 \cite{LH89}.

\subsubsection{Coupling constant} 
 Since the $\omega NN$ tensor coupling is generally 
 small (e.g. $\kappa_{\omega}=
 0.12$ in the vector dominance model), we take 
 $(g_{\omega},\kappa_{\omega}) = (10.1,0.0)$ as a typical strength.
 For the $\rho$ meson, we adopt the following two sets.
  
\vspace{1 cm}

\begin{center}

\begin{tabular}{|c|c|c|}\hline
 & set I  & set II   \\
\hline \hline
 $g_{\rho} $ & 2.63  & 2.72 \\
\hline
$\kappa_{\rho}$ & 6.0 & 3.7  
 \\ \hline 
\end{tabular} \\

\end{center}

\vspace{0.5cm}

\noindent
Table 2: Two sets of $\rho  NN$ coupling constants.
 $g_{\rho}$ ($\kappa_{\rho})$ denotes
 the vector (tensor) coupling.

\vspace{1 cm}

 Set I is obtained from the $N-N$ forward dispersion relation
 \cite{GK}.  The Bonn potential of the $N-N$ force gives 
 similar values with this set.
  Set II is obtained by the vector-meson dominance together  with the 
 $\rho$ universality \cite{sakurai}.
 A major difference between the two sets is the strength of the
 $\rho NN$ tensor coupling. (See ref.\cite{mac} for the
 detailed discussion on the vector-meson coupling constants.)
 In our calculations,  vertex form factors
 are not taken into account for simplicity.

\subsubsection{Numerical Results} 
As mentioned before, we will focus on the  time like region
 and consider the inverse propagator
\beq
\label{inverse}
D_T^{-1} (q_0,{\bf q}=0) = q^2 - m^2 +
  \Pi_T^D(q_0,{\bf q}=0) + \Pi_T^F(q_0^2) \ \ .
\eeq

 Let us define two kinds of masses $m_{re}^*$ (real mass) and 
 $m_{inv}^*$ (invariant mass).
 $m_{re}^*$ is defined as a lowest zero of $D_T^{-1}(q_0, 0)$.
  It is the quantity to be compared with that in the QCD sum rules.
 The invariant mass $m_{inv}^*$ is defined as a
  lowest zero of $D_T^{-1}$ with $\Pi_T^D$ neglected, in which case
  $D_T^{-1}$ is a function of $q^2$ only. 
 $m_{inv}^*$ here contains only the fluctuation of the Dirac sea
 by definition.  

 In Fig. 9, the effective masses of $\omega$ are shown together with
 $M^*_N/M_N$. The dashed line  denotes $m_{inv}^*/m$.
 One sees that
 Both $m_{re}^*$ and  $m_{inv}^*$  decrease at finite density,
 e.g. $m_{re}^*/m \simeq 0.8$ at $\rho=\rho_0$.
 The behavior of $m_{re,inv}^*$
   confirms the importance of the Dirac sea polarization
  found in previous studies \cite{FH,KS,TBA,CJ,PW,HS}.
  $m_{re}^* < m_{re}$ is caused by the fact that  
  the physical $\omega$ is more dressed by the $N \bar{N}$
 pairs in the medium since $M^*_N < M_N $ as we have
  discussed in sec.3.2 \cite{PW,HS}. 

\centerline{{\bf Fig.9}}
 
In Fig. 10  and Fig. 11, we have shown the effective masses
 of the $\rho$ meson with the parameter set I and set II, respectively.
 The strong $\rho NN$ tensor coupling plays a dominant role
  and gives 
 $m_{re}^*/m \simeq 0.6 - 0.7$ at $\rho =\rho_0$.  The 
 polarization of the Dirac sea is again the most important ingredient
 and the suppression of the wave-function renormalization factor $Z$
 is the main reason for the mass reduction. 

\centerline{{\bf Fig.10 and Fig.11}}

 It is in order here to make some remarks: 
 The  reduction of $m_{re}^*/m$ 
 is consistent with that in the QCD sum rules (eq.(7))
 for the $\omega$ meson, and even larger reduction is observed
 for the $\rho$-meson in QHD.
  One should, however, notice that we neglected  
 vertex form factors for the $\rho N N$ and $\omega N N$ couplings.
 Such  form factors will generally attenuate the magnitude of the
 mass shift of  $\rho$ and $\omega$.
 From Fig.9-11,  one
  also observes considerable non-linearity of $m_{re}^*$
 as a function of density, which is contrast to the linear
 dependence in eq.(7).

\section{$\phi$ meson in generalized QHD}

This section is partly based on the work in ref.\cite{kuwa}.
   The  $\phi$-meson
  is a  $\bar{s}s$  resonance in 	$J^P = 1^-$ channel
 with  a narrow width 
  ($m_{\phi} =1019.4$  MeV
 and $\Gamma_{\phi} = 4.4$ MeV). It  is a suitable probe of the
  partial restoration of chiral symmetry
   in hot/dense hadronic matter together with
    the $\rho$ and $\omega$ mesons \cite{finiteT,HL,theory}.
  In this section, we start with an effective hadronic model
 where the $\phi$-meson couples to  nucleon and 
 hyperons  ($B \equiv N, \Lambda^0, \Sigma^{\pm}, \Sigma^{0}$)
 with the vector coupling,
\beq
L_{int} = \sum_B g_{\phi B} \  \bar{B} \gamma_{\mu}B\  \phi^{\mu} .
\eeq
 $g_{\phi B}$ is the $\phi$-baryon coupling constant listed in
 Table 3.

\vspace{0.8cm}

\begin{center}

\begin{tabular}{|c|c|c|c|} \hline
Baryons     & $g_{\sigma B}$  & $g_{\omega B}$  &
 $g_{\phi B}$   \\ \hline \hline
 N          &8.7      &10.6    &4.2$^*$     \\ \hline
 $\Lambda$  &5.2      &6.9     &4.9         \\ \hline
 $\Sigma$   &5.2      &6.9     &4.9         \\ \hline
\end{tabular}\\

\vspace{0.5cm}

\end{center}

\noindent
Table 3: 
 $g_{\sigma B}$, $g_{\omega B}$ and $g_{\phi B}$ denote
 $\sigma$-$B$ scalar coupling,
  $\omega$-$B$ vector coupling,
  and 
  $\phi$-$B$ vector coupling, respectively.
  $g_{\sigma B}$ and  $g_{\omega B}$
 are taken from \cite{gled}. 

\vspace{0.8cm}

 Some remarks are in order here:\\
  (i) $\phi - \Lambda$ and $\phi - \Sigma$ couplings
 do not break the OZI rule, while
  the $\phi - N$ coupling is OZI violating.
 \\
 (ii)  $\Xi$ is neglected, since its effect to the 
 $\phi$ self-energy is doubly
 suppressed by the mass of $\Xi$ and by the OZI violation in
  $\phi-\Xi$
coupling.  \\
 (iii) If one relies on the quark counting rule \cite{QM},
  the $\phi$-hyperon couplings  are related to the
 $\omega$-hyperon  couplings as 
 $g_{\phi \Lambda (\phi \Sigma)} = 
g_{\omega \Lambda (\omega \Sigma)}/\sqrt{2}$ with
 $g_{\omega \Lambda (\omega \Sigma)}$ being determined by the
 fit of the hypernuclear levels \cite{gled}. This is assumed in Table 3. 
\\
  (iv) OZI violating $\phi$-nucleon coupling
  is not known experimentally.
  A study of the electromagnetic form-factors of the nucleon, however,  
 yields an upper bound of its strength \cite{jaffe}:
 $g_{\phi N}/g_{\omega N} <  0.4 $.

 For the non-strange nuclear matter,
  effects of the
 hyperons to the $\phi$-meson self-energy arise only through 
  hyperon$-$anti-hyperon loops,
  while the 
    nucleon contribution to the self energy 
  has both $N-\bar{N}$ loop
  and the Compton-type process.
  The one-loop self energy from hyperon and nucleon contributions
 reads
\beq
\label{polarization2}
\Pi_{\mu \nu}^B (q)
 =- {i \over (2\pi)^4}  \int d^4k \ g_{\phi B}^2\   
{\rm Tr}[\gamma_\mu G^H(k+q) \gamma_\nu G^H(k)]  \ \ ,
\eeq
where $G^H(k)$ denotes  baryon propagator in nuclear matter 
 given in (\ref{RHAS})  with $M_N^*$ replaced by $M_B^*$.

  Although one can calculate the density dependence
 of $M^*_B$ within QHD as in sec.3.1,
 we take here the following simple ansatz to illustrate the
  essential feature of the correlation between
   $M^*_B$ and $m_{\phi}^*$:
\beq
\label{baryon}
 g_{\sigma \Lambda (\sigma\Sigma)}/g_{\sigma N}
 & = & ( M_{\Lambda (\Sigma)} -M_{\Lambda (\Sigma)}^*)
/ (M_N - M_N^*), \\ 
\label{nucleon}
M_N^*/M_N & \simeq & 1 - 0.15 (\rho/\rho_0 ) .
\eeq
 Eq.(\ref{baryon}) is an 
 universal relation in QHD
 \cite{gled}
 and Eq.(\ref{nucleon}) is a simple parametrization 
 of  the effective nucleon mass which
   sometimes used in the literatures for $\rho < 2 \rho_0$ (see, e.g.
    \cite{CJ}).
	In Fig. 12, effective masses of $N$, $\Lambda$ and  $\Sigma$ 
 parametrized by eqs. (\ref{baryon},\ref{nucleon}) 
 are shown as a function of baryon density.

\vspace{0.2cm}

\centerline{{\bf Fig.12}}

\vspace{0.2cm}

 The $\phi$-meson mass $m^*_{\phi}$ in medium 
  is obtained as a  solution of the dispersion relation 
\beq 
\label{dispersion}
 \omega^2 - m_{\phi}^2 + \sum_B 
{\Pi}_B (\omega, {\bf 0})   =0,
\eeq
where
${\Pi}_B (\omega, {\bf 0}) \equiv -
 {\Pi}_B^{\mu\mu}(\omega, {\bf 0})/ 3\omega^2 $, and
 $m_{\phi}$ is the $\phi$-meson mass in the vacuum.
  ${\Pi}_B^{\mu \mu}$ is assumed to be a renormalized self-energy 
with the same subtraction procedure
   in the previous section.

 In Fig.13, 
 the ratio $m_{\phi}^*/m_{\phi}$ calculated 
  with only the hyperon-loops is shown by the solid line.
  $m_{\phi}^*$ decreases by $6\% $ at
 $\rho = \rho_0 $.
 Note that
  the Okubo-Zweig-Iizuka (OZI)
    rule is preserved for $\phi$-hyperon vertices, while it is
 violated in the
 self-energy  $ {\Pi}_B $.
  This is because the self-energy
  represents interaction
 of $\phi$ ($s\bar{s}$ pair)  with {\em non-strange} nuclear matter.
 Similar phenomena are known in two-step decay processes
 such as $\phi \rightarrow K\bar{K} \rightarrow \rho \pi$,
 $f' \rightarrow K\bar{K} \rightarrow \pi \pi$, and
$J/\psi \rightarrow D\bar{D} \rightarrow  \rho \pi$, where
 each vertex preserves the OZI rule while
  the whole amplitude does violate the rule  \cite{lipkin}.

\vspace{0.2cm}

\centerline{{\bf Fig.13, and  Fig.14}}

\vspace{0.2cm}

 The solid line in Fig.14 shows
 the ratio $m_{\phi}^*/m_{\phi}$ 
 calculated  with only the nucleon-loops. 
 Since $g_{\phi N}$ is not known experimentally,
 the result in this case has much uncertainty compared
 to the hyperon case.
  We have used $g_{\phi N}/g_{\omega N}
 = 0.32$ in Fig.14 which is close to the upper bound 
 given in Table 3: thus the
 resultant decrease of $m_{\phi}^*$ in Fig.14 should be considered as
  an upper
 limit.
 The negative mass shift in Fig.13 and Fig.14 is
 a direct consequence of the current conservation
 ($\partial_{\mu}(\bar{B}\gamma^{\mu}B)=0)$ and
 $M^*_B/M_B < 1$ as discussed in sec.3.2.
  This mechanism  is quite general and does not depend on the
 details of the interaction and on the virtual particles running 
 in the loop.
 
 To see the effect of the ultraviolet cutoff on the  finite
 part of the loop
 integral in (\ref{polarization2}),
  let us define  
  $ {\Pi}_B^{\mu \mu}(\omega, {\bf 0};\Lambda_{cut})
 \equiv \Pi_B^{\mu \mu}(\omega, {\bf 0};\Lambda_{cut}) -
 \Pi_B^{\mu \mu}(\omega, {\bf 0};\Lambda_{cut})\mid_{\rho =0}$ and use this
 in (\ref{dispersion}). We take  covariant cutoff for
 $\Lambda_{cut}$ for simplicity.
  When $\Lambda_{cut} \rightarrow \infty$, 
  $ {\Pi}_B(\omega, {\bf 0};\Lambda_{cut})$
 reduces to the renormalied ${\Pi}_B(\omega, {\bf 0})$.
 The dashed lines in Fig.13 and Fig.14
 are the results of such calculations for three cases, 
 $\Lambda_{cut}$ = 1, 2, 10  GeV.
  Although the cutoff dependence is not negligible,
 the qualitative picture we draw in the above is not affected.
 
We have considered only the nucleon and hyperon loops in the $\phi$
 self-energy.  Another possible contribution is the kaon-loop in medium.
 This was studied by Ko et al. \cite{theory} who
 found that the kaon-loop  has a tendency to decrease
  $m_{\phi}^*$ at low densities provided that the 
 effective kaon mass $\bar{m}_K^* = ( m^*_{K^{-}}+ m^*_{K^{+}})/2$
  decreases in medium.
  However,  whether $\bar{m}_K^*$ really decreases
 in nuclear matter or not  is still a controversial issue
  (see e.g. \cite{Yabu}).

\section{Experiments}

How one can detect the spectral change of vector mesons  in experiments?
 One of the promising ideas 
 is to use heavy nuclei and produce vector mesons in 
  $\gamma -A$ or $p-A$ reactions.
 Suppose one could create a vector meson at the center
 of a heavy nucleous. (It does not matter whether it is created  at the
 nuclear surface or at the center as far as the produced
 vector mesons run through the nucleous before the hadronic decay).
It is easy to see that the number of lepton pairs decaying inside the
 nucleous $N_{in}(l^+l^-)$ and that outside the nucleous
 $N_{out}(l^+l^-)$ are related as
\begin{eqnarray}
\label{NN}
{N_{in}(l^+l^-) \over N_{out}(l^+l^-)}  \sim  
{ 1 - e^{- \Gamma_{tot} R} \over e^{- \Gamma_{tot}R} } \ \ ,
\end{eqnarray}
where $\Gamma_{tot}$ denotes the total width of vector mesons
 ((1.3fm)$^{-1}$, (23fm)$^{-1}$ and (45fm)$^{-1}$ for $\rho$, $\omega$ and
 $\phi$, respectively) and $R$ being the nuclear radius.
 Eq.(\ref{NN}) shows that even the $\phi$ meson has considerable
 fraction of $N_{in}/N_{out}$ if the target nucleous is big enough.

 There exist already some proposals to look for the mass shift of 
 vector mesons in nuclear medium  \cite{SE}.
 One is by Shimizu et al. who  propose an experiment
 to  create $\rho$ and $\omega$
 in heavy nuclei using coherent photon - nucleus reaction and
  subsequently  detect the lepton pairs from $\rho$ and $\omega$.
 Enyo et al. propose to create $\phi$ meson in heavy nuclei
 using the proton-nucleus reaction and 
 to measure kaon pairs as well as the lepton pairs.
 By doing this, one can study not only the mass shift 
but also the change of the leptonic vs hadronic branching ratio
$r = \Gamma(\phi \rightarrow e^+ e^- ) /\Gamma (\phi \rightarrow K^+ K^- )$.
 Since $m_{\phi}$ is very close to $2m_{K}$ in the vacuum,
 any modification of the $\phi$-mass or the $K$-mass changes
 the ratio $r$ substantially as a function of mass number of 
 the target nucleous.
  Similar kinds of experiments are also planned
 at CEBAF and GSI.

 There are also on-going  heavy ion experiments at SPS (CERN) and AGS (BNL)
 where  high density matter is likely to be formed.
 In particular, CERES/NA45 and HELIOS-3 at CERN
   reported enhancement of the
  lepton pairs below the $\rho$ resonance \cite{ceres,helios},
  which may not be explained
 by the conventional sources of lepton pairs.
  E859 at BNL-AGS reported a possible spectral change of the
 $\phi$-peak in $K^+K^-$ spectrum \cite{ags}.
 If these effects are real, low mass enhancement  of the lepton pair
  spectrum
 expected by the mass shift of the vector mesons could be a
  possible explanation \cite{ceres2} (see also \cite{ceres3}
 for another explanation). 
  In nuclear collisions at  higher energies (RHIC and LHC),
  hot hadronic matter or possibly the quark-gluon-plasma 
 with low baryon density are expected to be formed.  In such cases,
 double $\phi$-peak structure proposed by Asakawa and Ko \cite{theory}
 as well as  the spectral change of $\rho$ and $\omega$ \cite{HKL}
will be a distinct signal of the chiral restoration in QCD.

\section{Concluding remarks}

The spectral change of the elementary excitations in medium
 is an exciting new possibility in QCD.  
 By studying such phenomenon, one can learn the structure
 of the hadrons and the QCD ground state at finite $(T, \rho$)
 simultaneously.   Theoretical approaches such as the QCD sum rules
 and the hadronic effective theories 
 predict that the  light vector mesons ($\rho$, $\omega$ and $\phi$) 
 are sensitive to the partial restoration of chiral symmetry
 in hot/dense medium.  These mesons are good 
 probes experimentally, since they decay into lepton pairs which penetrate
 the hadronic medium without loosing much information.
  Thus, the lepton pair spectroscopy in QCD will tell us a 
 lot about the detailed structure of the hot/dense matter, which
 is quite similar to the soft-mode spectroscopy
 by the photon and neutron scattering experiments in solid state physics.

\vspace{1cm}

This work was supported by the Grants-in-Aids of the Japanese Ministry
 of Education, Science and Culture (No. 06102004). T. H. thanks
 Institute for Nuclear Theory at the University of
 Washington for its hospitality and the
 Department of Energy for partial support during
 the completion of this work. 

\newpage
\centerline{{\bf Figure Captions}}
\begin{description} 
\item [Fig.1]
The light quark condensates in N=Z nuclear matter in the
 linear density approximation.
  Theoretical uncertainty of the $\pi N$ sigma term is
  taken into account. We take $y = 0.12$  for 
  the OZI breaking parameter, where 
 $y \equiv 2\la \bar{s}s \ra_N/ \la \bar{u}u + \bar{d} d \ra_N$
 with $\la \cdot \ra_N$ being the nucleon matrix element.

\item [Fig.2]
 Borel curve for the $\rho(\omega)$ meson mass.
 Solid, dashed and dash-dotted lines
 correspond to $\rho/\rho_0$ = 0, 1.0 and 2.0 respectively. $S_0^*(\rho)$ 
 determined by the
 Borel stability method at each density is also shown in $\GeV^2$ unit.
The Borel window is chosen to be $0.41 \GeV^2 < M^2< 1.30 \GeV^2$. 

\item [Fig.3]
 Masses of $\rho, \omega$ and $\phi$ mesons 
 in nuclear matter predicted in the QCD sum rules.
  The hatched region shows  theoretical uncertainty.

 \item [Fig.4]  Feynman diagram of the relativistic Hartree 
 approximation (RHA).
 The wavy (dashed) line is the vector (scalar) meson propagator.
The double (single) solid line  denotes 
 the Hartree (free) propagator of the nucleon.  
  \item [Fig.5]  An illustration of  the divergent parts of $\Sigma_s$.
 The dashed line is the scaler meson propagator and 
 the double (single) solid line  denotes 
 the Hartree (free) propagator of the nucleon.  
 \item [Fig.6]  An illustration of
 the contribution from  counter terms.
 Crosses denote the insertion of the counter terms
  with coefficients $\alpha_1-\alpha_4$.

\item [Fig.7] (a) The Dyson equation for the 
 vector meson propagator with self-energy 
 $\Pi$.
 The double (single) wavy line is the full (free) vector-meson propagator.\\
 (b) Vector-meson self-energy  in the one-loop approximation.
 The double solid line  denotes the nucleon propagator in the
  Hartree approximation.

\item [Fig.8]
Effective $\omega$ meson mass as a function of  the baryon density
  using three different subtraction procedures for meson self-energy.
The solid, dashed and dash-dotted lines  are obtained by the
 subtraction schemes given 
in ref.\cite{PW},\cite{HS} and \cite{KS}, respectively.
 Triangle, diamond and square denote the values at
  $\rho = \rho_0$ in each subtraction scheme.

\item [Fig.9] Effective masses of the nucleon and the $\omega$ meson
as a function of the baryon density. $m^*_{re}$ ($m$) denotes
 the real mass in the medium (the mass in vacuum). 
  The dashed line corresponds to the
invariant mass in the medium. 
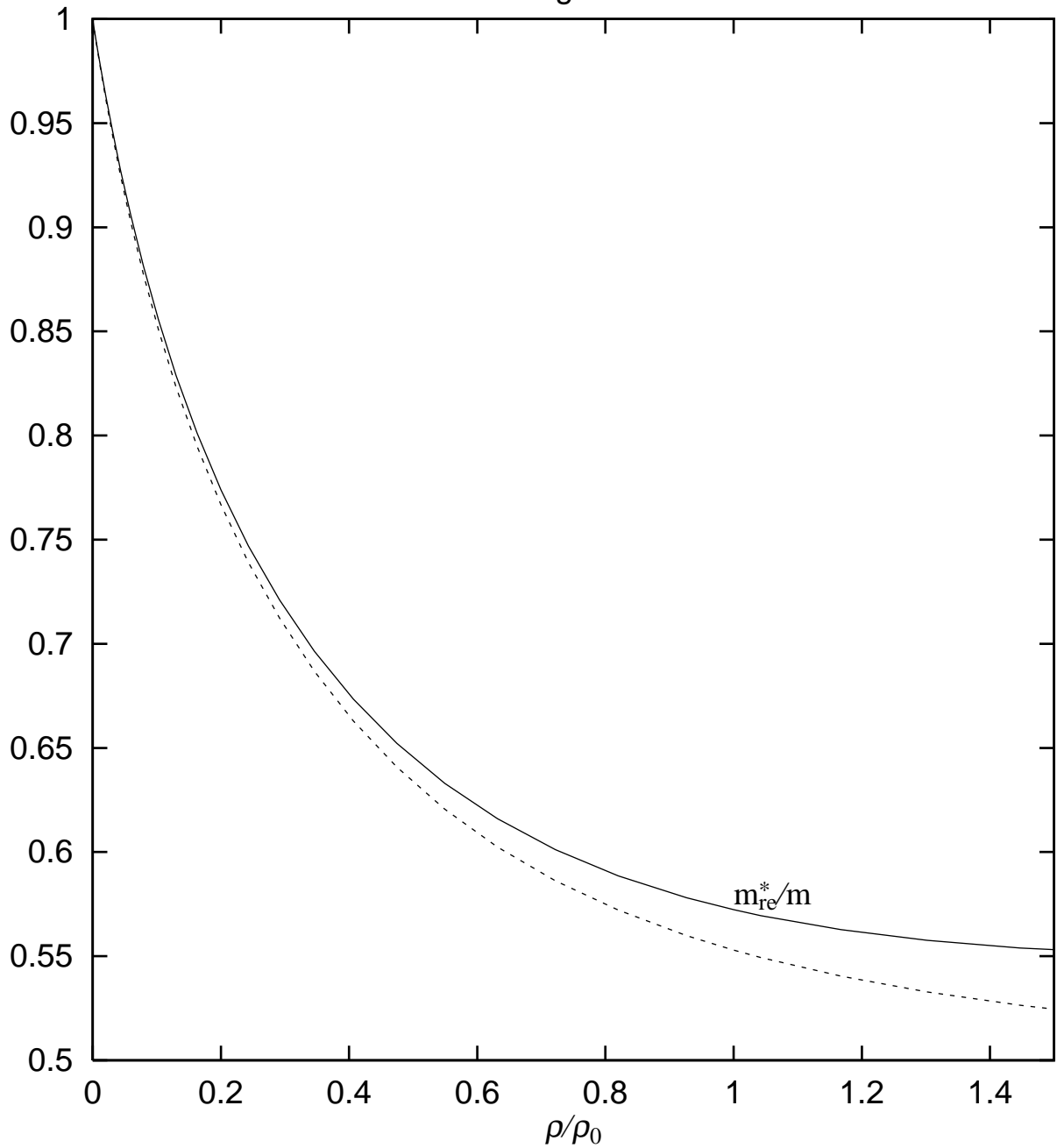
\item [Fig.10]
 Real and invariant masses of the $\rho$-meson in the parameter
 set I, ($g_\rho, \kappa_\rho$)=(2.63,6.0). The dashed line corresponds
 to $m^*_{inv}/m$. 
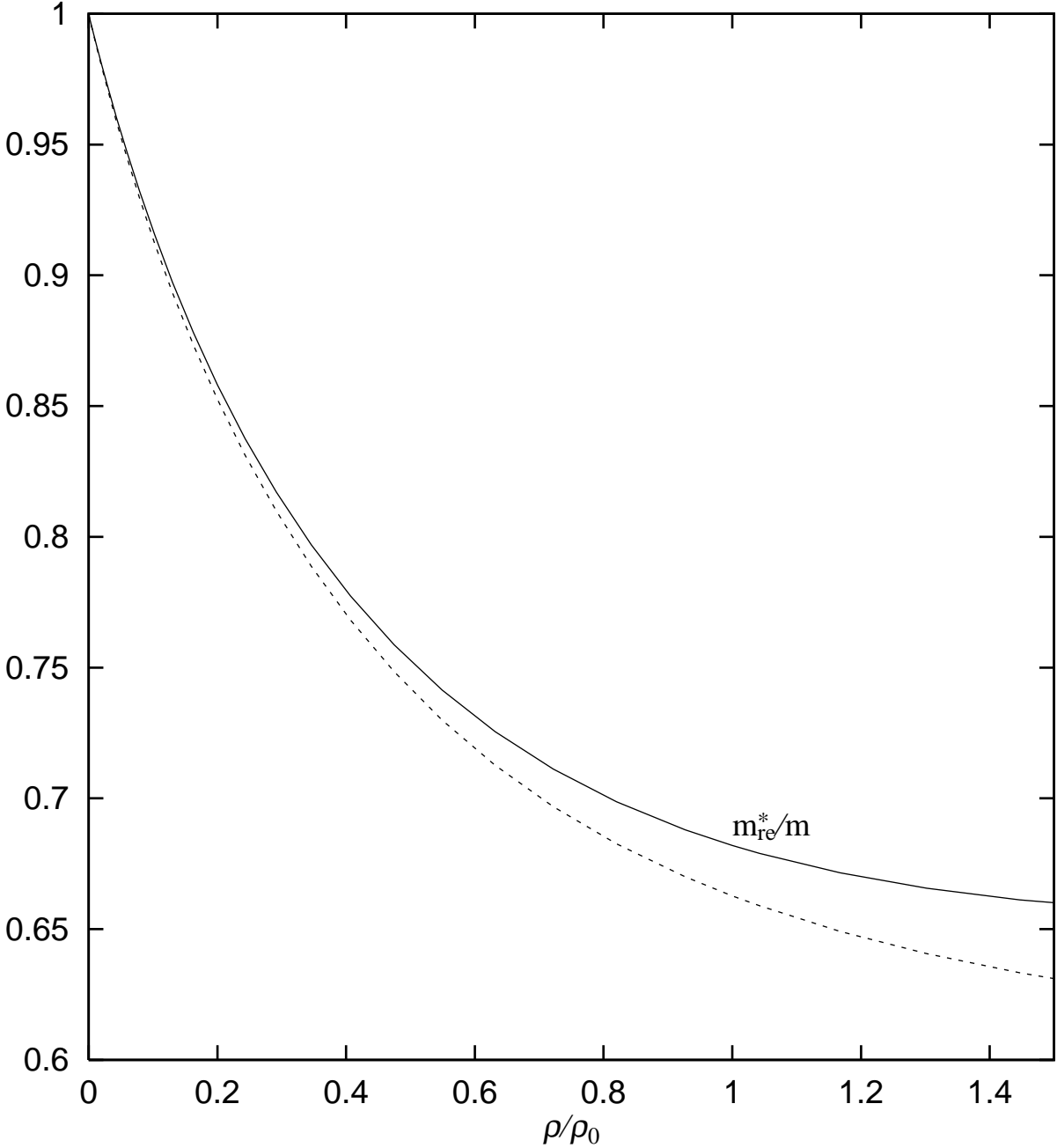
\item [Fig.11]
 Same  quantities with Fig.10 in the parameter set II,
  ($g_\rho, \kappa_\rho$)=(2.72,3.7). The dashed line corresponds
 to $m^*_{inv}/m$. 

\item [Fig.12] 
 Ratio of the baryon mass in matter $M_{B}^*$ and that in vacuum 
 $M_{B}$ ($B=N, \Lambda, \Sigma$) as a function of  $\rho/\rho_0$.
  This figure is taken from ref.\cite{kuwa}.

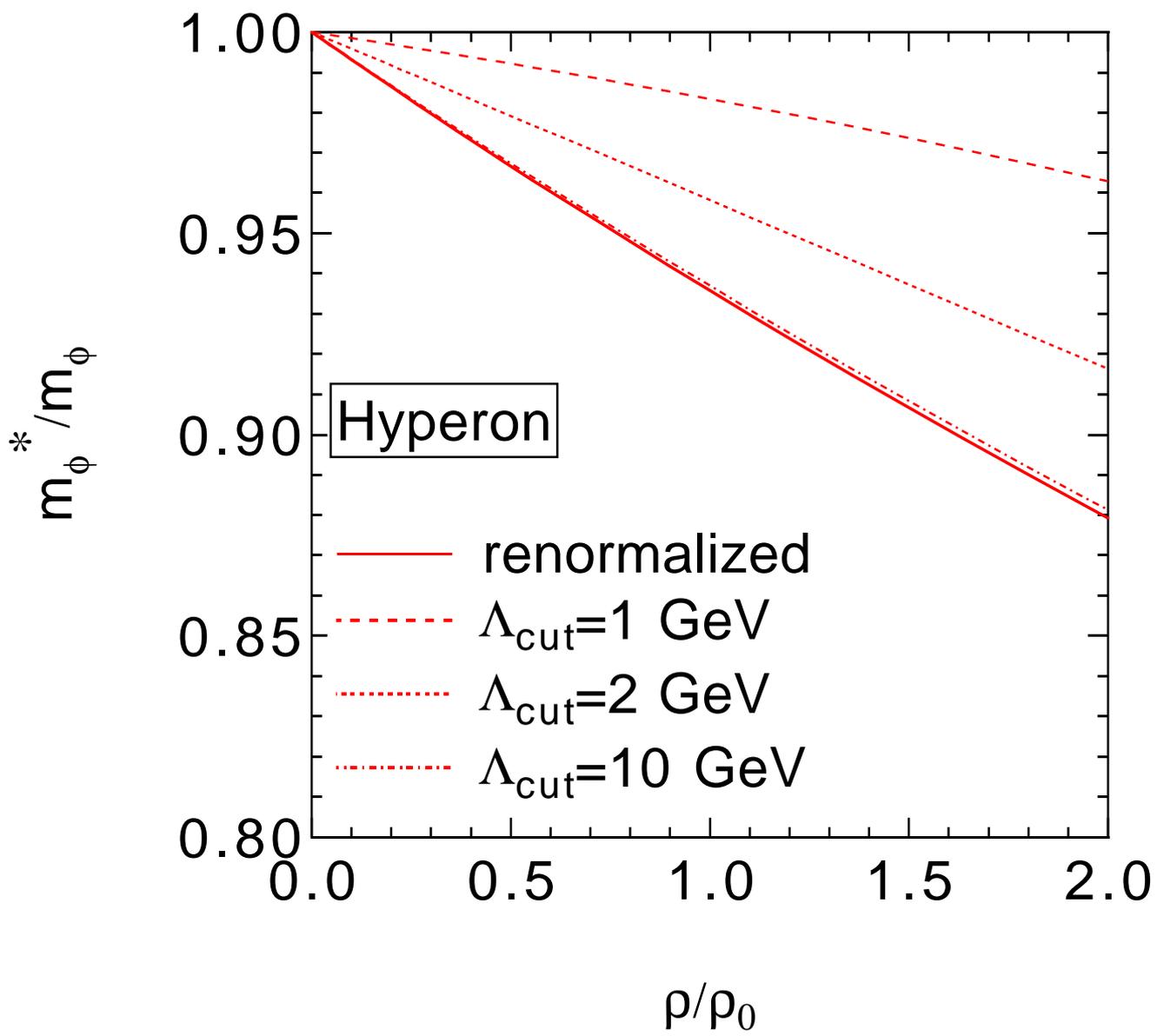
\item [Fig.13]
 Ratio of the $\phi$-meson mass in matter $m_{\phi}^*$ and that
 in the vacuum $m_{\phi}$ as a function of $\rho/\rho_0$.
 Only hyperon contributions are included in the $\phi$-meson self-energy.
   This figure is taken from ref.\cite{kuwa}.

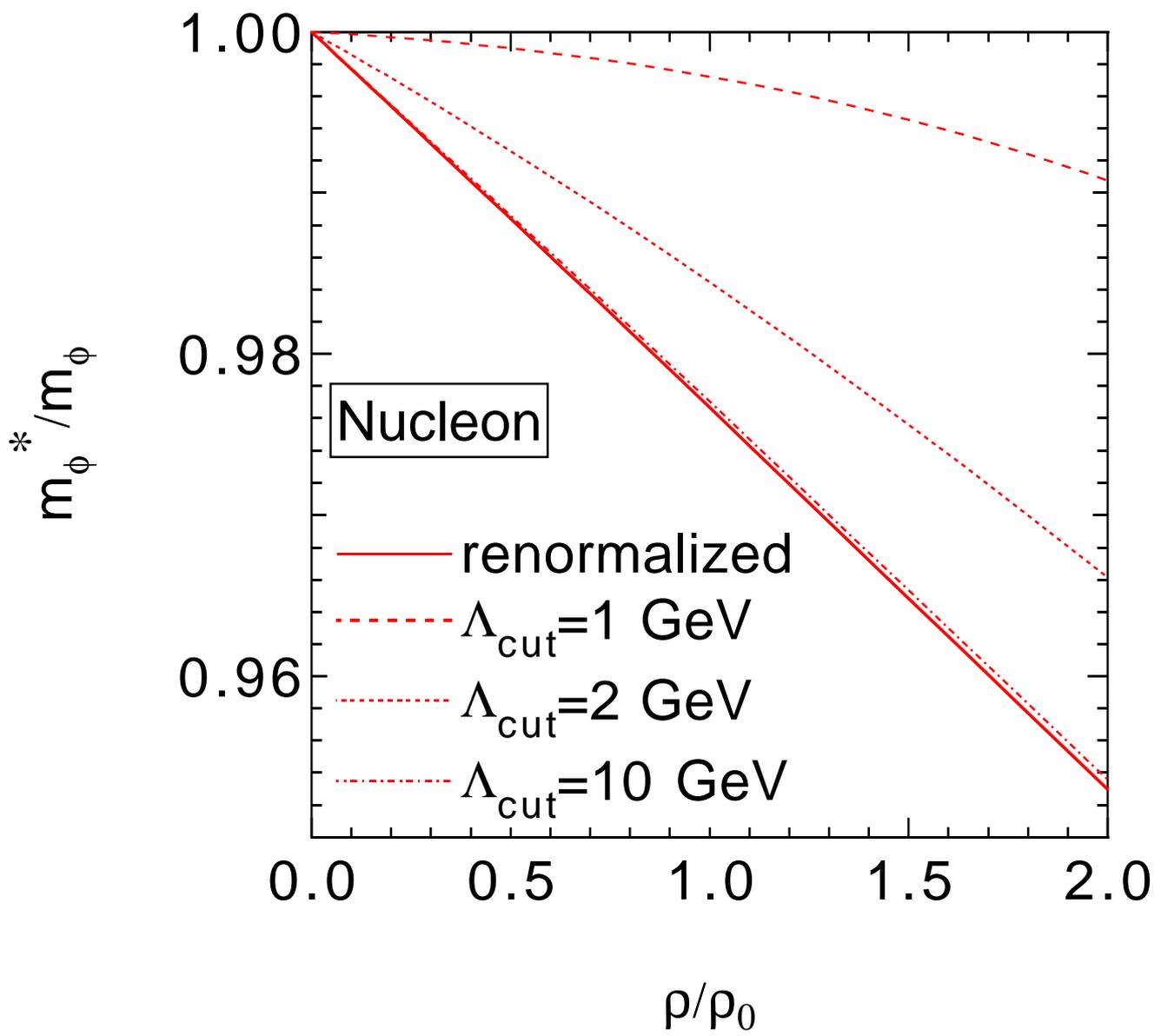
\item [Fig.14] Same with Fig.13 except that
 only the nucleon contribution is included.
 This figure is taken from ref.\cite{kuwa}.
 
\end{description}

\end{document}